\documentclass[a4paper]{article}

\usepackage{INTERSPEECH2021}
\usepackage{multirow}
\usepackage{color}
\usepackage{amssymb}
\usepackage{bbding}

\usepackage{graphicx} 
\usepackage{float} 
\usepackage{subfigure} 
\usepackage{array}
\usepackage{booktabs}

\title{Continual Learning for Fake Audio Detection}
\name{Haoxin Ma$^{1,2}$, Jiangyan Yi$^1$, Jianhua Tao$^{1,2,3}$, Ye Bai$^{1,2}$, Zhengkun Tian$^{1,2}$, Chenglong Wang$^1$}
\address{$^1$NLPR, Institute of Automation, Chinese Academy of Sciences, China\\
$^2$School of Artificial Intelligence, University of Chinese Academy of Sciences, China \\
$^3$CAS Center for Excellence in Brain Science and Intelligence Technology}

\email{mahaoxin2019@ia.ac.cn, \{jiangyan.yi, jhtao\}@nlpr.ia.ac.cn}

\begin{document}
\maketitle
\begin{abstract}
  Fake audio attack becomes a major threat to the speaker verification system. Although current detection approaches have achieved promising results on dataset-specific scenarios, they encounter difficulties on unseen spoofing data. Fine-tuning and retraining from scratch have been applied to incorporate new data. However, fine-tuning leads to performance degradation on previous data. Retraining takes a lot of time and computation resources. Besides, previous data are unavailable due to privacy in some situations. To solve the above problems, this paper proposes detecting fake without forgetting, a continual-learning-based method, to make the model learn new spoofing attacks incrementally. A knowledge distillation loss is introduced to loss function to preserve the memory of original model. Supposing the distribution of genuine voice is consistent among different scenarios, an extra embedding similarity loss is used as another constraint to further do a positive sample alignment.
  Experiments are conducted on the ASVspoof2019 dataset. The results show that our proposed method outperforms fine-tuning by the relative reduction of average equal error rate up to \(81.62\%\).


\end{abstract}

\section{Introduction}


Fake audio detection has attracted increasing attention since the organization of a series of automatic speaker verification spoofing and countermeasures challenge (ASVspoof) \cite{todisco2019asvspoof,kinnunen2017asvspoof,wu2015asvspoof}.
Plenty of approaches are proposed to alleviate the threat of voice spoofing attacks. The studies on fake audio detection are usually carried out from two aspects. The first is robust acoustic features based on signal processing methods \cite{patel2015combining, todisco2016new, das2020assessing}. The second is effective classifiers based on neural networks \cite{lavrentyeva2017audio, Lavrentyeva2019, lei2020siamese, von2020multi}.
Despite the impressive results that these methods have achieved, we can find that the performance of these models degrades when unseen spoofing attacks occur. For example, in term of the median equal error rate (EER) of the top-10 performing submitted systems in ASVspoof2019 challenge, the median EER under seen attack A16 is $0.02\%$ while it goes dramatically up to $15.93\%$ under unseen attack A17 \cite{todisco2019asvspoof}. This phenomenon reveals the weakness of current countermeasures and urges people to look for ways to improve model's the performance on unseen attacks.


Some studies pay attention to fake audio detection on unseen datasets. Monteiro et al. \cite{monteiro2020ensemble} proposed a model ensemble approach to train three models jointly, which outperforms the model trained with mixed data directly. Wang et al. \cite{wang2020dual} proposed a dual-adversarial domain adaptation framework to solve out-of-domain dataset problem.
However, these methods require both original and new data. With the growth of voice conversion and speech synthesis technologies as well as the emergence of advanced reply devices, new spoofing attacks keep emerging rapidly. It's storage-consuming to do model ensemble and time-consuming to train model with mixed data involving all the data sources. Besides, for the consideration of privacy, the access to old data may not be allowed in some special cases. Although fine-tuning the trained model is another way to improve the performance on unseen spoofing attack \cite{wang2020dual}, it will lead to a decrease on the previous spoofing knowledge.
Therefore, it's necessary to make our detection model have the ability to continually learn new knowledge over time meanwhile maintain the previously learned knowledge.

\begin{figure}[t] 
\vspace{-10pt}
\centering 
\includegraphics[width=0.7\linewidth]{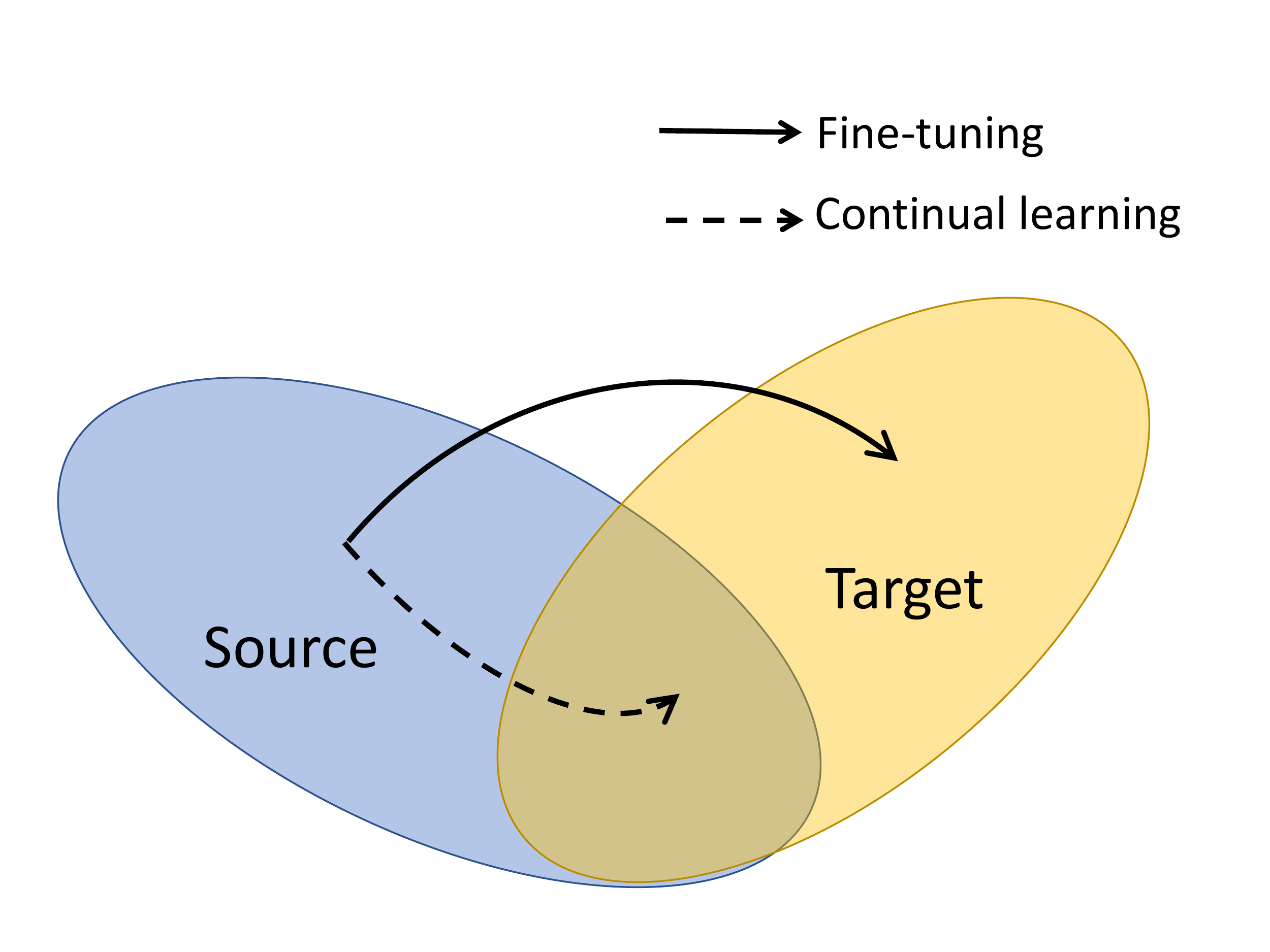} 
\vspace{-5pt}
\caption{Difference between continual learning methods and fine-tuning.
Blue and yellow region is the parameter configuration region of source and target models correspondingly. If using the fine-tuning strategy (solid arrow), the results of parameter optimization will easily fall into the region deviation from the source, while continual learning methods (dotted arrow) can learn the parameters in the overlapping region.} 
\label{ewc} 
\vspace{-5pt}
\end{figure}

Continual learning becomes a hot issue in recent years \cite{kirkpatrick2017overcoming,rebuffi2017icarl,van2020brain} and has been applied in the field of computer vision \cite{castro2018end, lee2020residual} and speech recognition \cite{sadhu2020continual, houston2020continual} already. It aims to overcome the catastrophic forgetting problem existing in fine-tuning, i.e., the model forgets previously learned information after trained on new information. The difference between continual learning and fine-tuning is elaborated in Figure \ref{ewc}.

In this work, we propose a continual learning method, named detecting fake without forgetting (DFWF), to learn new spoofing attacks incrementally. It is inspired by learning without forgetting (LwF) but takes the characteristics of fake audio detection into account. When model is trained with new unseen data, genuine speech is more consistent in different scenarios while spoofing speech varies widely. Thus, we add an extra positive sample alignment (PSA) constraint on genuine speech.
DFWF can help model mitigate the forgetting of past knowledge meanwhile focus more on the invariance of genuine speech.
Our proposed method is investigated on several sequential training tasks across different spoofing types on ASVspoof2019 dataset. The results show that DFWF approach can reduce the forgetting of past knowledge and its performance is much better than fine-tuning.
To our best knowledge, this is the first study to apply continual learning to the fake audio detection task to solve the problem that current models have difficulties in distinguishing unseen spoofing audio.

The rest of this paper is organized as follows: Section 2 illustrates our proposed method DFWF from its two components: LwF and PSA. Experiments, results and discussions are reported in Section 3, 4 and 5, respectively. Finally, the paper is concluded in Section 6.

\section{Proposed methods}
\subsection{Proposed detecting fake without forgetting}

Usually, there are three major approaches for continual learning: (1) Regularization approaches \cite{li2017learning, kirkpatrick2017overcoming} add hand-craft regularization term in loss function to constrain the learning process. (2) Replay approaches \cite{rebuffi2017icarl,shin2017continual} record some examples in the buffer and replay experience when training new task. (3) Dynamic architecture approaches \cite{rusu2016progressive, parisi2017lifelong} dynamically change the structure of neural networks according to the new task. We focus on regularization approaches in the study.

Since retraining with mixed data costs a lot of time and fine-tuning leads to catastrophic forgetting, we propose DFWF method to make a trade off between time resource and performance.
It doesn't store old data but can ensure the previous information is remembered.
When training, two constrains named learning without forgetting (LwF) and positive sample alignment (PSA) are added to the original loss $\mathcal{L}_{original}$. Thus, the total loss ${\mathcal{L}_{total}}$ becomes:
\vspace{-2pt}
\begin{equation}
    {\mathcal{L}_{total}} =  \mathcal{L}_{original} + \alpha {\mathcal{L}_{LwF}} + \beta {\mathcal{L}_{PSA}}
  \label{totalloss}
\end{equation}
where $\alpha$ and $\beta$ are hyper-parameters to control the importance of respective terms.
Usually, cross entropy loss is used as $\mathcal{L}_{original}$ to train the model. The details of LwF loss ${\mathcal{L}_{LwF}}$ and PSA loss ${\mathcal{L}_{PSA}}$ are explained in the following subsections.

\subsection{Learning without forgetting}

LwF\cite{li2017learning} is introduced to DFWF method. When training new task, LwF uses knowledge distillation loss to make the output probabilities of new data on current network close to the output probabilities of the same data on original network. Thus, it can also be seen as an extension of student-teacher knowledge distillation model \cite{hinton2015distilling} which is widely adopted in speech recognition \cite{chebotar2016distilling, shi2019knowledge}. LwF is achieved by including an additional knowledge distillation loss $\mathcal{L}_{LwF}$:

\vspace{-5pt}
\begin{equation}
  \mathcal{L}_{{\rm{LwF}}} = - \sum\limits_{i = 1}^n {{y'}_{old}^{(i)}}\log {{{y'}_{new}^{(i)}}}
\end{equation}
where $n$ is the number of labels. The label includes genuine and fake in this case. $i$ means the i-th class. ${y'}_{old}^{(i)}$ and ${y'}_{new}^{(i)}$ are modified probabilities of original model's output ${y}_{old}^{(i)}$ and current model's output ${y}_{new}^{(i)}$. Specifically, they are calculated as follows:
\vspace{-5pt}
\begin{equation}
    {y'}_{old}^{(i)} = \frac{{{{(y_{old}^{(i)})}^{1/T}}}}{{\sum\nolimits_j {{{(y_{old}^{(j)})}^{1/T}}} }},
    \qquad
    {y'}_{new}^{(i)} = \frac{{{{( y_{new}^{(i)})}^{1/T}}}}{{\sum\nolimits_j {{{( y_{new}^{(j)})}^{1/T}}} }}
  \label{LwF2}
\end{equation}
where $T$ is a hyper-parameter. Setting $T > 1$ increases the weight of smaller values in logits and encourage the network to produce a relative similar contributions between the old and the new data\cite{hinton2015distilling}.

\subsection{Positive sample alignment}
In the ``genuine or fake" classification task, the feature distribution of genuine speech is more consistent compared with the feature distribution of various types of fake audio. In a new spoofing scenario, there exists a gap between previous spoofing data and new spoofing data. But the feature of genuine speech in different situations tend to be similar. When meeting new data, the model should focus more on the genuine samples to succeed more knowledge of the original model for genuine samples. Considering this characteristic of fake audio detection, PSA is proposed to constrain the feature distribution of genuine speech among different data resource.
Cosine distance is used to evaluate the similarity of positive (genuine) embeddings between the current model and the original model.
The proposed PSA loss ${\mathcal{L}_{PSA}}$ is denoted as:
\vspace{-5pt}
\begin{equation}
    {\mathcal{L}_{PSA}} = \frac{1}{{{N_P}}}\sum\limits_{k = 1}^{{N_P}} {\frac{{y_k^{+}\hat y_k^{+}}}{{\left\| {y_k^{+}} \right\| \cdot \left\| {\hat y_k^{+}} \right\|}}}
  \label{psa1}
\end{equation}
where ${N_P}$ is the number of genuine speech. ${y_k^{+}}$ and $\hat y_k^{+}$ are the embedding vectors (the outputs of the network after removing the last full connection layer) of the $k$-th genuine speech extracted by the original model and the current model, respectively. $\left\| \cdot \right\|$ means the norm of the vector. Thus, PSA loss equals to $\cos {\theta _{y_k^{+}\hat y_k^{+}}}$, the cosine distance of genuine speech embeddings between the previous and current model.

\section{Experiments}
\subsection{Dataset}
All experients are conducted on ASVspoof 2019 dataset \cite{wang2020asvspoof} which has two subset:  logical access (LA) and physical access (PA). The LA subset includes 19 types spoofing attacks made by voice conversion or speech synthesis. The PA subset contains 27 different acoustic and 9 different replay configurations. To further evaluate the performance of our proposed methods on a series of unseen spoofing attacks, we select four spoofing attacks with big difference in LA and PA subset separately. More specifically, in LA subset, type A13, A17, A10, A19 are chosen. Since only LA evaluation set contains these types, we further redivide them into training, development and evaluation sets. In PA subset, type bbbBB, cccCC, aaaAC, abcAA are chosen. Because all of these PA types occur in the training, development and evaluation sets, their partitioning scheme are according to their divisions in the PA subset. The detailed statistics of the datasets are presented in Table \ref{tab:dataset}.

\begin{table}[th]
\vspace{-5pt}
\setlength\tabcolsep{3.5pt}
  \caption{ Number of utterances in each section of the dataset}
  \vspace{-5pt}
  \label{tab:dataset}
  \centering
  \begin{tabular}{ cccc }
    \toprule
    \multirow{2}*{\textbf{Dataset}} & \multicolumn{3}{c}{\textbf{Number of utterances}} \\
    \cline{2-4}
    ~ & \textbf{Training} & \textbf{Development}  & \textbf{Evaluation}\\
    \midrule
    ASVspoof2019LA             & 25380 & 24986  & 71933    \\
    ASVspoof2019PA             & 54000  & 33534  & 153522   \\
    LA-A13                        &  2250  & 1250 & 1352     \\
    LA-A17                    & 2250 & 1250  &1321         \\
    LA-A10                 & 3414 & 1500 & 1500         \\
    LA-A19                   & 3414 & 1500 &1500         \\
    PA-bbbBB                    & 356  & 278     & 1078          \\
    PA-cccCC                    & 472  & 336     & 1380          \\
    PA-aaaAC                    & 604  & 402     & 1654          \\
    PA-abcAA                    & 352  & 276     & 916         \\
    \bottomrule
  \end{tabular}
\vspace{-10pt}
\end{table}

\subsection{Evaluation metrics}
The performance of our proposed methods is evaluated via equal error rate (EER) and average EER (AvgEER).
The lower the EER value, the higher the accuracy of the model. AvgEER is introduced when the model is evaluated in sequential trainings. It calculates the arithmetic mean of EER among all the spoofing types.
\vspace{-5pt}
\begin{equation}
    AvgEER = \frac{1}{n}\sum\limits_{i = 1}^n {{{EER}_i}}
  \label{avgeer}
\end{equation}
\vspace{-5pt}

where $n$ is the number of spoofing types in testing.${EER}_i$ is the EER of trained model tested on the $i$-th spoofing types.



\subsection{Experimental setup}
We extract 60-dimensional linear frequency cepstral coefficients (LFCC) \cite{sahidullah2015comparison} of speech as low-level input feature to the neural network. The extraction way is the same as the ASVspoof2019 offical baseline system \cite{wang2020asvspoof}. The window length is set to 25 ms. The number of FFT bins is set to 512. The number of filters is set to 20. To avoid the influence of short audio duration, we duplicate the short audio along time-axis up to 320 frames. If the audio is longer than that set length, we just randomly take a segment of the set length.

As for the network, we adopt light convolutional neural network (LCNN) which is the same architecture as Lavrentyeva et al. \cite{lavrentyeva2019stc} to generate 80-dimensional high-level embeddings. Then a fully connection layer projects the embeddings to a binary classification.

To evaluate the performance of DFWF, we compare it with two methods: fine-tuning and multi-condition (retraining) training. Fine-tuning refers to train a new model with new data by taking the weights of previous model as initialization. Multi-condition training refers to retrain the model by incorporating new date with old data.
It's worth noting that the EER of multi-condition training is considered to be the lower bound to our continual-learning-based method DFWF \cite{parisi2019continual}.
In addition, we did an ablation study to illustrate the relative contribution of LwF and PSA to the final performance of DFWF. In order to cover comprehensive cases of sequential unseen spoofing attacks, we designed the following 4 types of sequential training settings:

\begin{enumerate}
\item $PA \rightarrow LA$ : This setting is to evaluate the performance of DFWF method across the spoofing types with obvious gap.
\item $LA \rightarrow PA$: This setting is the contrast to $PA \rightarrow LA$ to evaluate the influnence of transfer order.
\item $A13\rightarrow A17\rightarrow A10\rightarrow A19$: This setting is to evaluate the performance of DFWF method across the spoofing types with relatively subtle difference. These are all LA attacks. In the order $A13\rightarrow A17\rightarrow A10\rightarrow A19$, we train the new model with new spoofing data on the basis of the model trained at the last step.
\item $bbbBB\rightarrow cccCC\rightarrow aaaAC \rightarrow abcAA$: This setting is to evaluate the performance of DFWF method across different PA attacks.
\end{enumerate}

We name the model trained at the first step as base model. Then, new model is trained with different method in the following steps. At each step, the model is trained for 100 epochs by Adam optimiser with the learning rate 0.0001. For DFWF method, hyper-parameter $T$ is experimentally set to 2. We set weighting coefficients $\alpha$ and $\beta$ ranging from $0.0$ to $2.0$ to explore the optimal parameter for each corresponding task.

\section{Results and Analysis}

\subsection{Catastrophic forgetting in fake audio detection}

We do several experiments to fine-tune the model with unseen spoofing attacks and find that there exists catastrophic forgetting phenomena. Considering the result of ``$LA \rightarrow PA$" experiment shown in the first two row of Table \ref{tab:lapa}, the EER of base model increases significantly when tested on PA. After fine-tuning on PA, the fine-tuned model can perform well on PA, but its EER under LA subset increases sharply from $4.42\%$ to $19.78\%$. This is because of the distribution mismatch between LA attacks and PA attacks. When fine-tuned with new data, the model will easily deviate from the previous optimal parameters, resulting in performance decrease on old data.
Thus, it's necessary to apply our continual learning method DFWF to unseen spoofing data scenario. Fine-tuning is used as the baseline of the following experiments.

\subsection{The effectiveness of our proposed DFWD}

We compare our proposed DFWF method with fine-tuning and multi-condition training in 4 sequential training experiments.

\begin{table}[t]
\vspace{-5pt}
\setlength\tabcolsep{3pt}
  \caption{Performance comparison: results show the of EER\% on different models for training sequence ``$LA \rightarrow PA$" and ``$PA \rightarrow LA$". Base model is trained on one dataset with random initialization. Fine-tuning (FT) model and DFWF model are trained on the other dataset on the basis of base model. Multi-condition (MC) training model is trained on the mixture of LA and PA subsets.}
  \vspace{-5pt}
  \label{tab:lapa}
  \centering
  \begin{tabular}{ cccccccc }
    \toprule
    \multirow{2}*{Model} & \multicolumn{3}{c}{{$LA \rightarrow PA$}} & \multicolumn{3}{c}{{$PA \rightarrow LA$}} \\
    \cmidrule(r){2-4}  \cmidrule(r){5-7}
    ~ & LA & PA & Avg & PA & LA & Avg \\
    \midrule
    Base     & 4.42 &  28.02  & 16.22
                        & 5.47 & 22.31 & 13.89\\
    FT  & 19.78 &  6.74 & 13.26
                         & 21.22  & 4.61 & 12.92\\
    DFWF (Ours)    & 7.74  &  10.76 & 9.25
                        & 8.85	& 12.05 &  10.45 \\
    MC     &  5.92 &  6.29 & 6.11
                                & 6.29  & 5.92  &6.11\\
    \bottomrule
  \end{tabular}
\end{table}

\begin{figure}[t]

  \centering
  \subfigure[LA attacks]{
    \label{Fig.la4}
    \includegraphics[width=0.45\linewidth]{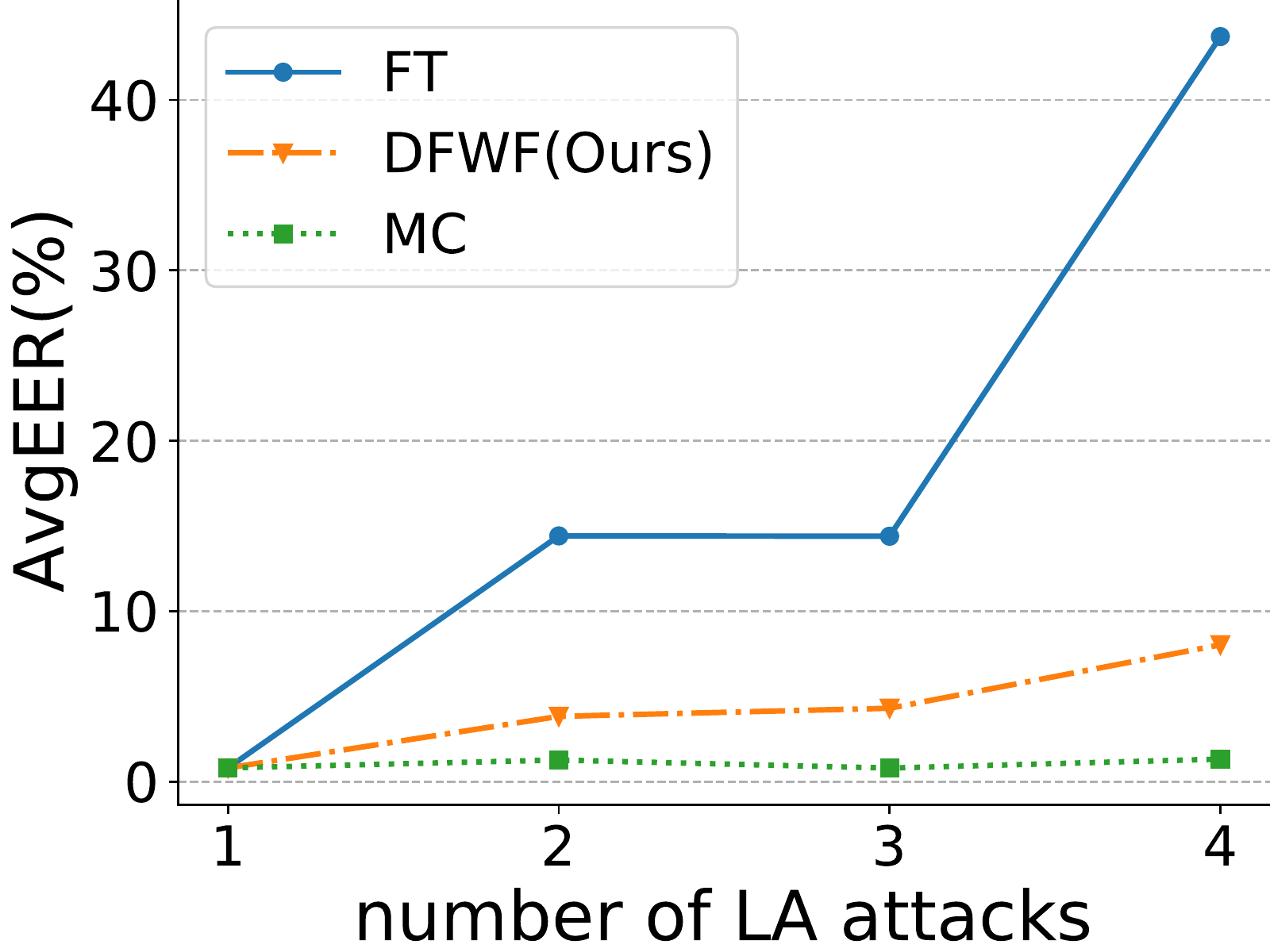}}
    \subfigure[PA attacks]{
    \label{Fig.pa4}
    \includegraphics[width=0.45\linewidth]{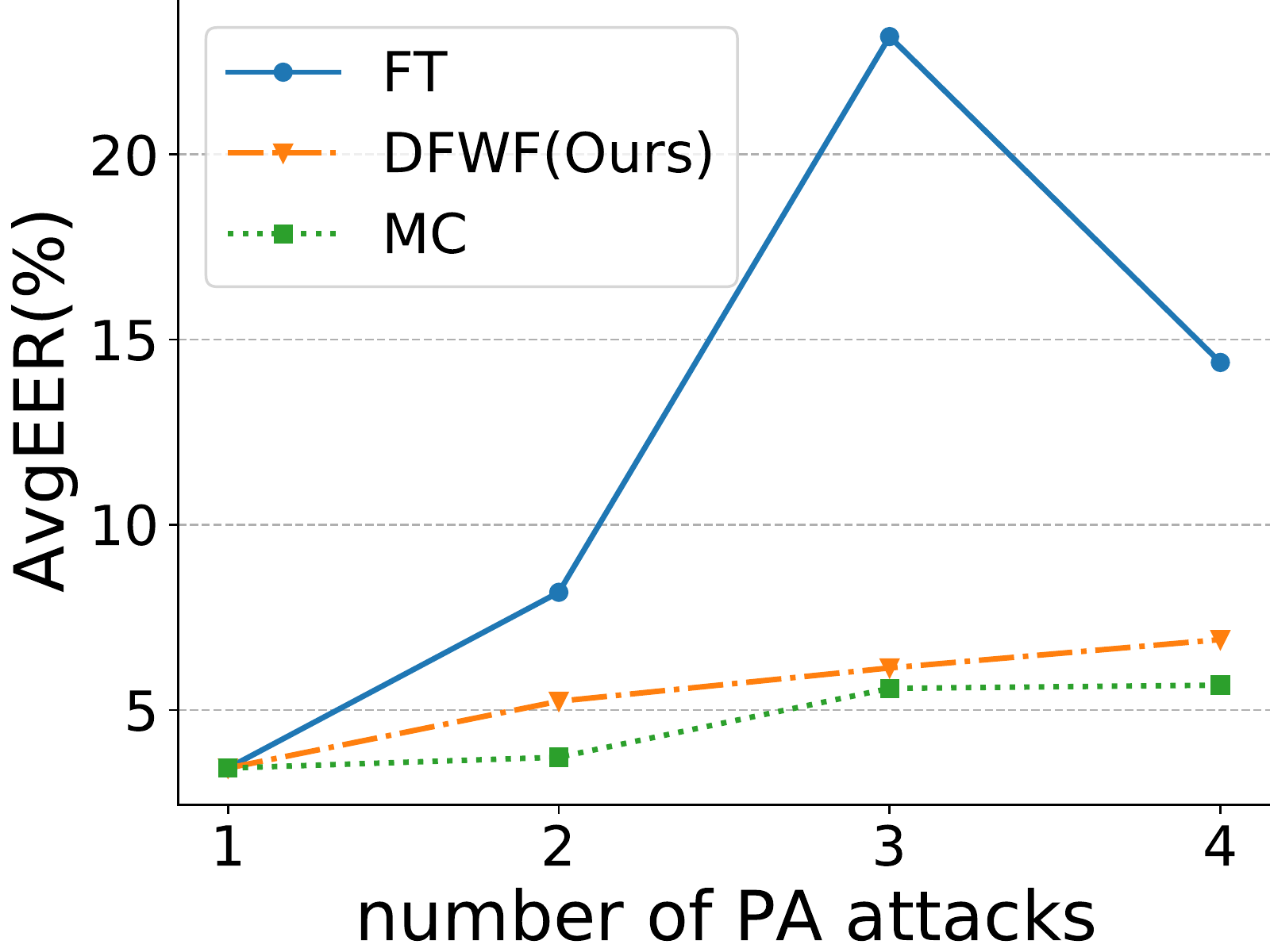}}
  \caption{Performance comparison under incremental increase of spoofing types: results show the AvgEER\% over the sequential training of LA attacks ($A13\rightarrow A17\rightarrow A10\rightarrow A19$) and PA attacks ($bbbBB\rightarrow cccCC\rightarrow aaaAC \rightarrow abcAA$). FT means fine-tuning model. MC means multi-condition training model. AvgEER when the number of attacks equals to $i$ means the average performance on current and previous data at the $i$-th training step. }
  \label{fig:4types}
\vspace{-5pt}
\end{figure}

Table \ref{tab:lapa} illustrates the result of sequential training between LA and PA attacks. The AvgEER of DFWF achieves a relative reduction of $30.29\%$ in ``$LA \rightarrow PA$" task, and a relative reduction of $19.08\%$ in ``$PA \rightarrow LA$" task compared to fine-tuning baseline.
As this baseline does not consider the previous data in its training procedure, it has the advantage of achieving better performance on current data but damages the performance on old data.
Besides, the transfer order can influence the performance of DFWF method. The AvgEER of ``$LA \rightarrow PA$" task is better than ``$LA \rightarrow PA$" task. It can be explained by the distribution of LA attacks and PA attacks provided by the publisher of ASVspoof2019 dataset \cite{wang2020asvspoof}, different LA attacks can form relatively clear clustering while different PA attacks are completely overlapping. The model trained on LA subset can been seen as a better pre-trained model, providing a set of well-performed model parameters.

The advantage of DFWF is more pronounced in the scenarios where unseen spoofing attacks occur gradually .
Figure \ref{fig:4types} shows the results in sequential trainings of LA and PA attacks, respectively. In sequential training``$A13\rightarrow A17\rightarrow A10\rightarrow A19$", the performance of fine-tuned model becomes worse as the number of spoofing types increases. When the types increase to 4, the AvgEER of fine-tuned model goes dramatically from $0.81\%$ to $43.72\%$, while the AvgEER of DFWF model remains to $8.03\%$. That is to say, DFWF achieves a relative reduction of $81.63\%$ compared to fine-tuning in this task. In the sequential training ``$bbbBB\rightarrow cccCC\rightarrow aaaAC \rightarrow abcAA$" , DFWF model achieves a AvgEER of $6.90\%$ fairly close to the multi-condition training model ($5.67\%$) at the last step. Besides, we see a obvious fluctuation in the trend of AvgEER with fine-tuning strategy. This can be explained by the existence of similarity among different spoofing types. If the feature new spoofing is similar to some of the previous spoofing types, training with current data can help to maintain some of the previous knowledge.



\subsection{Ablation Study}
\begin{table}[t]
 \vspace{-10pt}
\setlength\tabcolsep{5.5pt}
  \caption{Comparison of AvgEER\% in different combinations of LwF and PSA in 4 sequential trainings. T1 is the training ``LA $\rightarrow$ PA". T2 is the training ``PA $\rightarrow$ LA". T3 is the training `` $A13\rightarrow A17\rightarrow A10\rightarrow A19$". T4 is the training ``$bbbBB\rightarrow cccCC\rightarrow aaaAC \rightarrow abcAA$". }
  \vspace{-5pt}
  \label{ablation study}
  \centering
  \begin{tabular}{ cccccc }
    \toprule
    + LwF & + PSA   & T1 & T2 & T3 & T4 \\
    \midrule
   \XSolidBrush  & \XSolidBrush   &13.26& 16.01 & 43.72 & 14.38\\
   \Checkmark  & \XSolidBrush   & 12.28 & 11.68  &  11.52 &  7.44   \\
     \XSolidBrush &  \Checkmark   & 9.64 & 11.79 & 12.28  & 8.48 \\
    \Checkmark  & \Checkmark  & 9.25 & 10.45 & 8.03  &  6.90 \\
    \bottomrule
  \end{tabular}
 \vspace{-5pt}
\end{table}

We carry out an ablation study to evaluate the impact of two components of our method. The first factor LwF is the constraint on the output distribution of model for all input data. The second factor PSA is the constraint on embeddings of genuine speech input. Both of them are used for preserving key parameters consistent with the original model. Table \ref{ablation study} shows the AvgEER achieved in all combinations of these two components. Training without neither LwF nor PSA equals to fine-tuning, and training with both LwF and PSA equals to DFWF.
In training, we adopt the optional hyper-parameters for each setting. The results show that both two components can improve the performance in sequential training. When applied together, DFWF obtains the best AvgEER.

\section{Discussions}

\begin{figure}[t]
\vspace{-10pt}
  \centering
  \subfigure[LCNN-BASE]{
    \label{Fig.1}
    \includegraphics[width=0.45\linewidth]{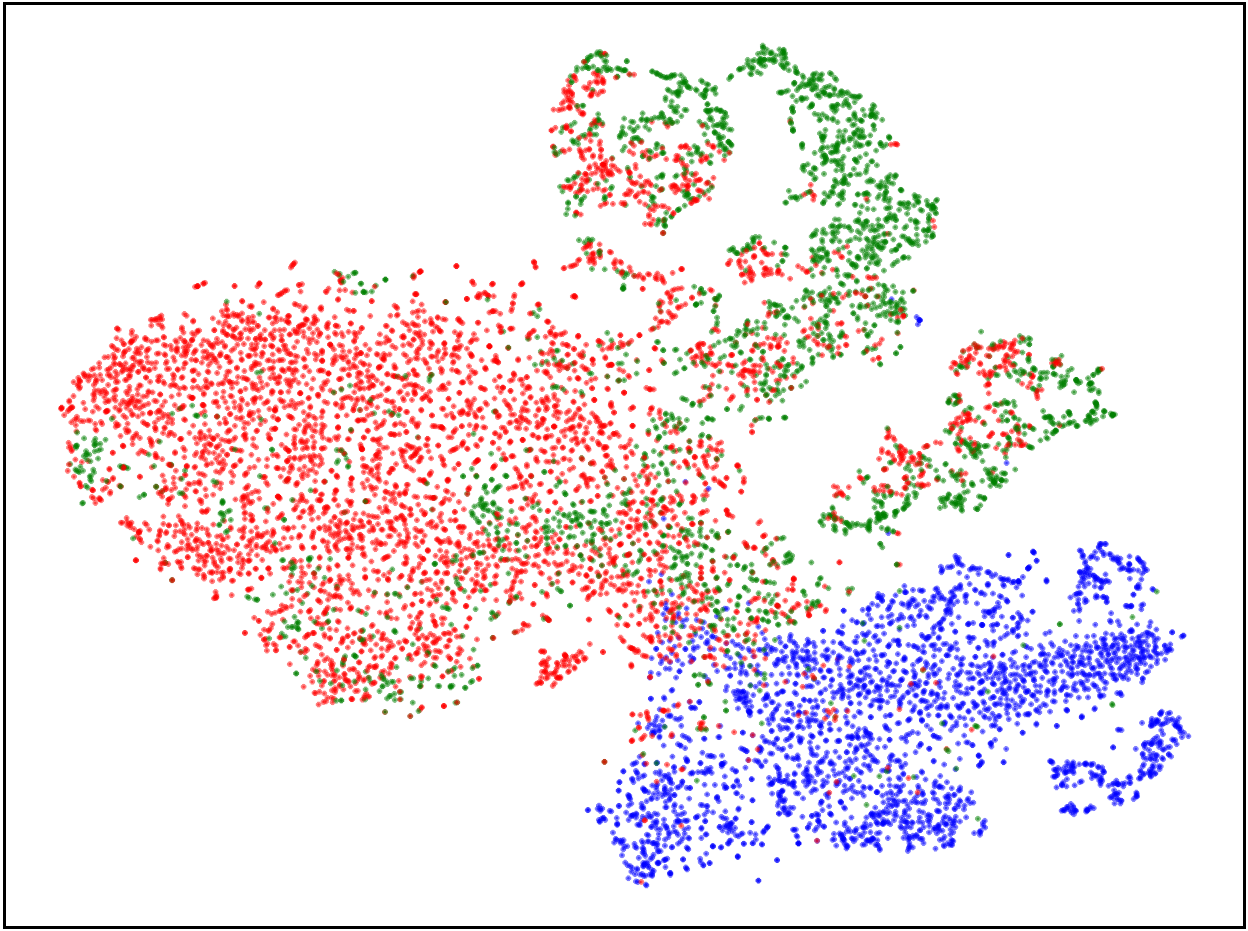}}
    \subfigure[LCNN-FT]{
    \label{Fig.2}
    \includegraphics[width=0.45\linewidth]{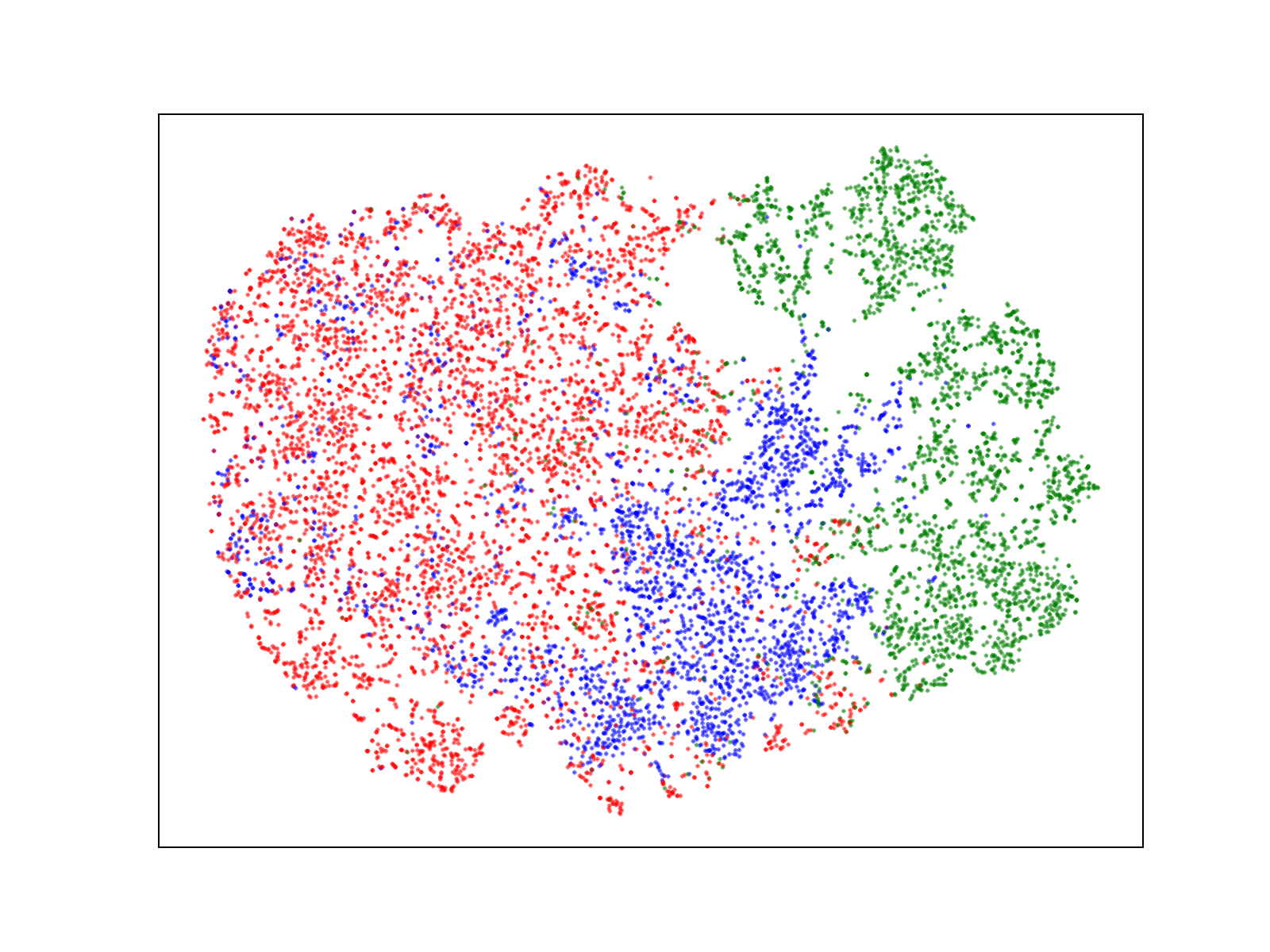}}
    \subfigure[LCNN-DFWF (Ours)]{
    \label{Fig.3}
    \includegraphics[width=0.45\linewidth]{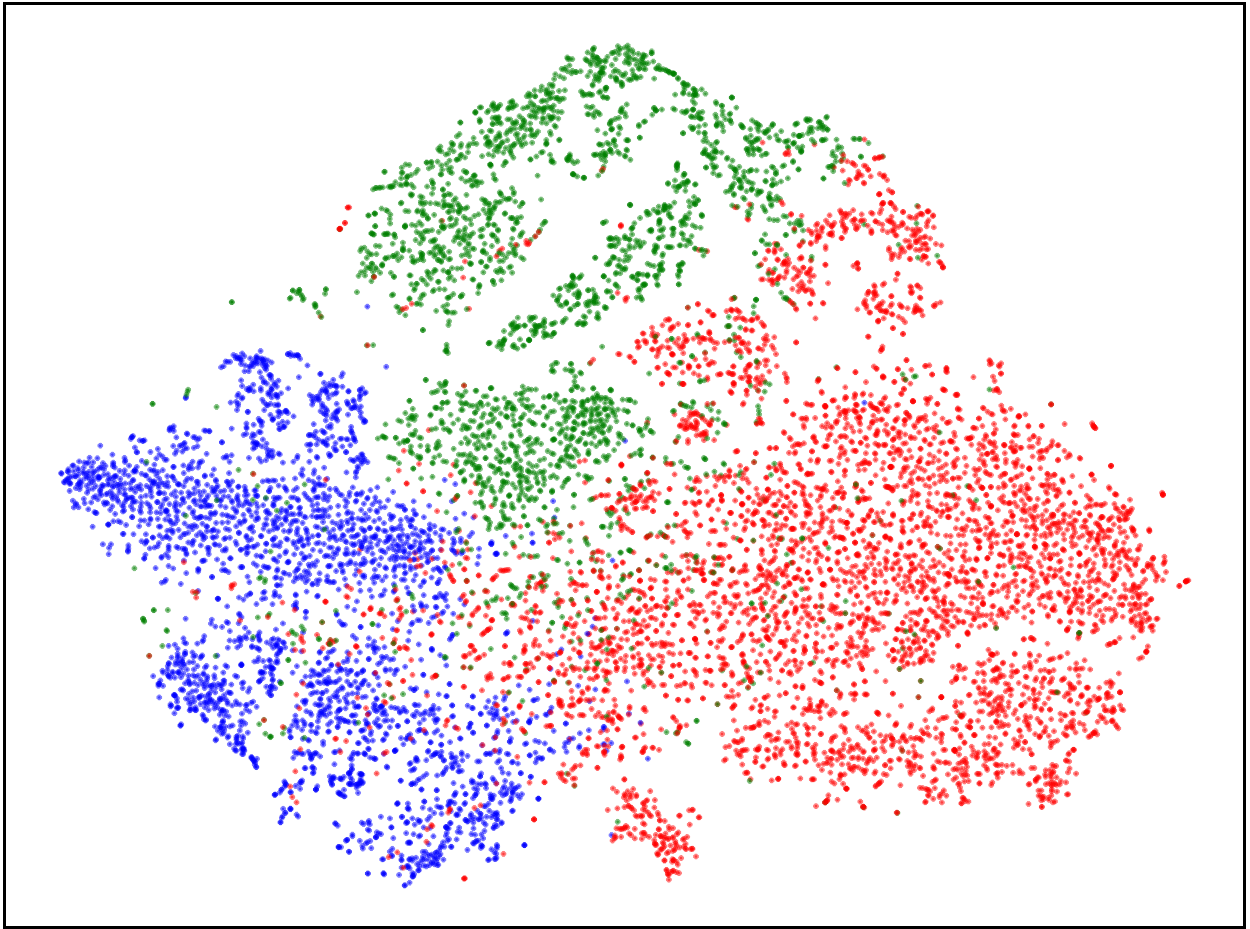}}
    \subfigure[LCNN-MC]{
    \label{Fig.4}
    \includegraphics[width=0.45\linewidth]{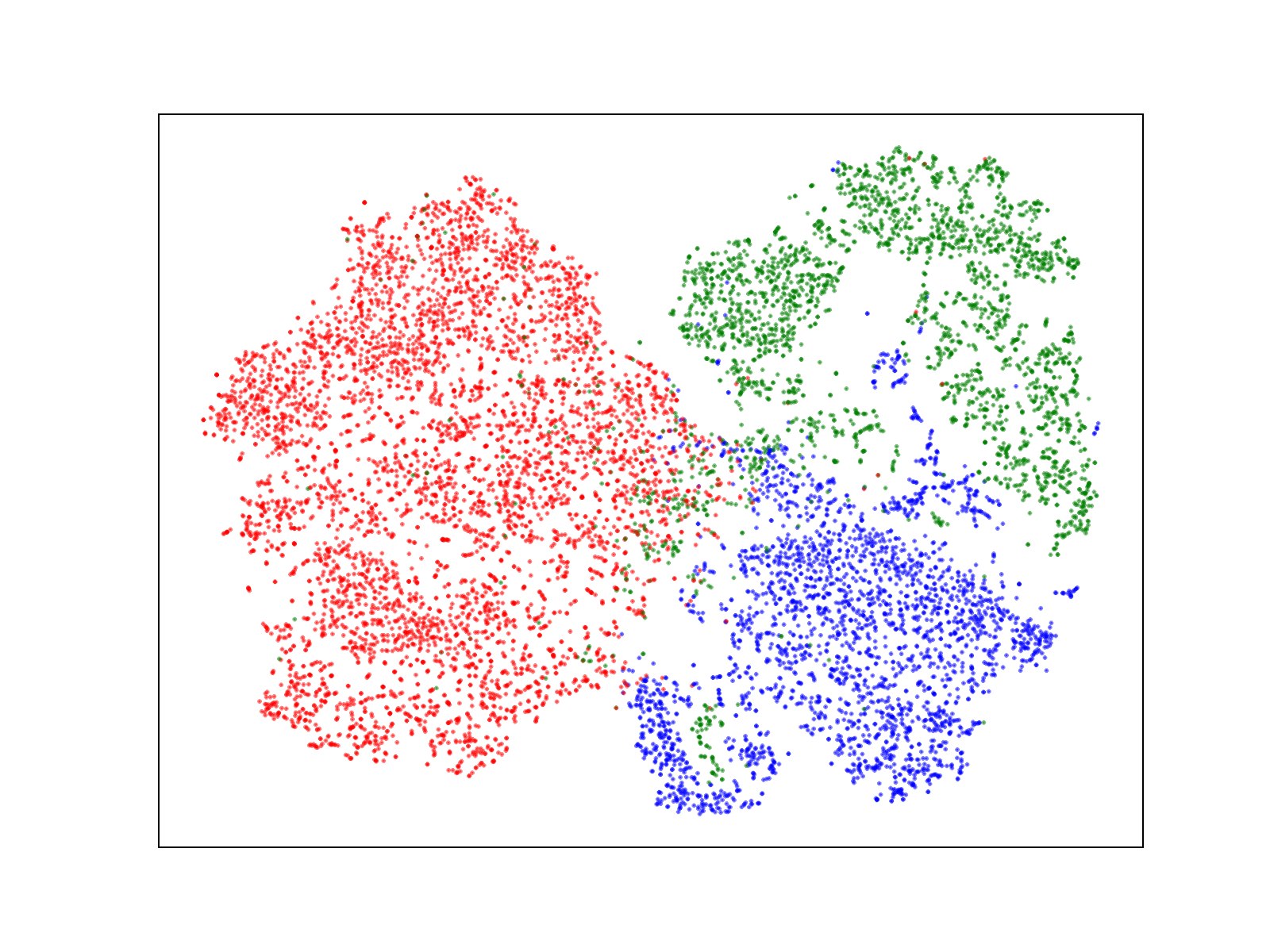}}
  \vspace{-10pt}
  \caption{The t-SNE visualization of all training data embeddings that are extracted by different models. LCNN-BASE is trained only on LA subset. LCNN-FT is finetuned on PA subset based on LCNN-BASE. LCNN-DFWF is trained on PA by the propoesd DFWF method based on LCNN-BASE. LCNN-MC is train on mixed data of LA and PA subsets. The color red represents genuine speech. The color blue represents spoofing speech of LA attack. The color green represents spoofing speech of PA attack.}
  \label{fig:embedding}
\vspace{-10pt}
\end{figure}

From the results, we find that our proposed DFWF doesn't outperform multi-condition training. The reason is that all the data are available to multi-condition model but the previous data is inaccessible to our DFWF systems. In theory, the results of multi-condition model is the upper limit of the performance of continual learning systems\cite{van2008visualizing}.

Besides, to better understand the mechanism of our proposed DFWF method, we plot the t-SNE figures to visualize the embeddings of genuine and spoofing speech \cite{van2008visualizing}. Figure \ref{fig:embedding} shows the embedding distribution of all training data in ``$LA \rightarrow PA$" sequential training.
LCNN-BASE can not distinguish the spoofing speech of PA attack (green dots) since there is an obvious mixture of PA spoofing speech and genuine speech (red dots). After fine-tuned on PA subset, LCNN-FT can distinguish PA attacks but the previous LA attacks (blue dots) are scattered among genuine speech. These illustrate the dilemma of fine-tuning that the fine-tuned model tends to forget the previous knowledegs.
However, after applied DFWF, LCNN-DFWF can distinguish both LA attacks and PA attacks much more better than the LCNN-FT, and the distribution of the data is very similar to that in LCNN-MC.



\section{Conclusions}
In this paper, we proposed a continual-learning-based method DFWF to solve the unseen spoofing attacks problem in fake audio detection. On the basis of LwF, PSA prevents the embedding of genuine speech from deviating from the original distribution, which further preserves the model's detection capability on previous data.
This method can not only save the time and computing resource but also mitigate the catastrophic forgetting problem. In some cases ( i.e., $bbbBB\rightarrow cccCC\rightarrow aaaAC \rightarrow abcAA$), it even performs similarly to multi-condition training but it's faster to train without the need to get access to the previous data.
Experiments conducted on 4 setting of sequential training show the effectiveness of DFWF method to learn new spoofing attacks incrementally, particularly as spoofing data becomes more and more.
More advanced continual learning methods are worth exploring in fake audio detection in the future.


\bibliographystyle{IEEEtran}

\bibliography{mybib}


\end{document}